# The Quantum Field Of A Magnet Shown By A Nanomagnetic Ferrolens

*Emmanouil Markoulakis[a]\*, Antonios Konstantaras[a], Emmanuel Antonidakis[a]*

[a]Technological Educational Institute of Crete, ComputerTechnology, Informatics & Electronic Devices Laboratory, Romanou 3, Chania, 73133, Greece

A B S T R A C T

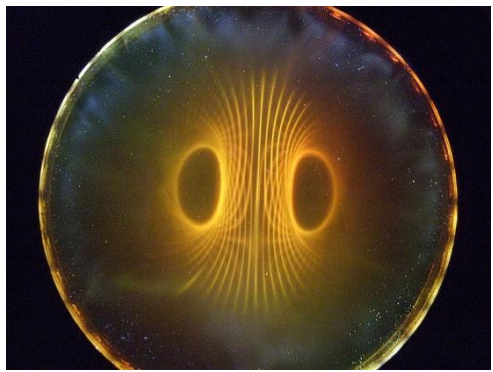

**Graphical Abstract**

It has been more than two hundred years since the first iron filings experiment, showing us the 2D macroscopic magnetic imprint of the field of a permanent magnet. However, latest developments in modern nanomagnetic passive direct observation devices reveal in real-time and color a more intriguing 3D dynamic and detailed image of the field of a magnet, with surprising new findings, that can change our perspective for dipole magnetism forever and lead to new research.

This research is a continuation of our previous work, *"Markoulakis, E., Rigakis, I., Chatzakis, J., Konstantaras, A., Antonidakis, E. Real time visualization of dynamic magnetic fields with a nanomagnetic ferrolens(2018) Journal of Magnetism and Magnetic Materials, 451, pp. 741-748.DOI: 10.1016/j.jmmm.2017.12.023"* that is using a ferrolens apparatus for showing the dynamic magnetic field on a transmitting radio antenna, while this time the magnetostatic fields were under our scope and examined with the aid of the ferrolens. We are presenting experimental and photographical evidence, demonstrating the true complex 3D Euclidian geometry of the quantum field of permanent magnets that have never been seen before and the classic iron filings experiment, apart of its 2D limitations, fails to depict. An analysis of why and what these iron filings inherent limitations are, giving us an incomplete and also in some degree misguiding image of the magnetic field of a magnet is carried out, whereas, as we prove the ferrolens is free of these limitations and its far more advanced visualization capabilities is allowing it to show the quantum image with depth of field information, of the dipole field of a permanent magnet.

For the first time the domain wall (i.e. Bloch or Neel wall) region of the field of a magnet is clearly made visible by the ferrolens along with what phenomenon is actually taking place there, leading to the inescapable conclusion, novel observation and experimental evidence that the field of any dipole magnet actually consists of two distinct and separate toroidal shaped 3D magnetic bubbles, each located at either side of the dipole around the exact spatial regions where the two poles of the magnet reside.

## 1. Introduction

We are using the same ferrolens device, introduced and described by us in our previous research [1] to show the actual 3D geometry of the dipole field of permanent magnets and as such for any other dipole static magnetic field, since the geometry of a magnet's dipole magnetic field does not depend or change with the shape of the magnet. Their field is uniform and remains geometrically the same for all dipole magnets. Very little research has been carried out so far about the topic of 3D field geometry of permanent magnets and they all rely and are based on the old, 2D iron filings macroscopic experiment [2] imprint of the field shown in **fig.1** .

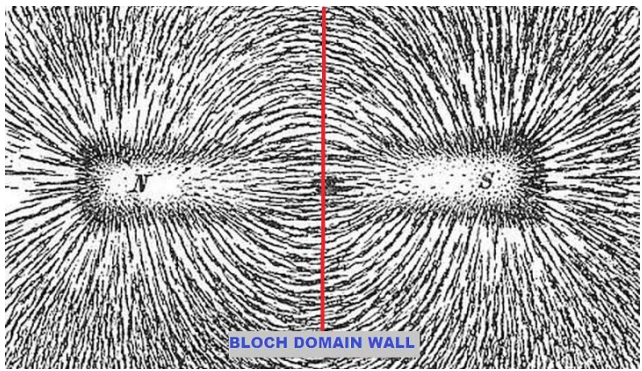

**Fig. 1. Classic iron filings experiment with N-S poles and Bloch domain wall indicated.**

However this picture, of the iron filings, besides, their apparent 2D limitation for depicting the field, is due to their strong ferromagnetism, size, and their magnetic interference, lacking in the *fine tuning*, sensitivity and resolution required to depict the very important details of a static magnetic dipole field. Moreover, they are not suitable for use in the actual 3D visualization of the field.

For example the very low near zero, magnetic reluctance [3] of iron filings will cause them to actually behave more like a compass needle. They always orient themselves relative to their position towards the highest potential regions of the dipole static field namely the two poles of the magnet. Therefore the totally miss to show what is actually happening to the magnetic flux tangent to the field force vectors, near the Bloch (or preferably Neel) domain wall [3] region at the middle of a permanent magnet **fig.2,** a region of diminishing magnetic field strength *(important clarification, from here on when we referring to the Bloch region we are referring to the field area of a magnet which is near and around to its Bloch domain wall fig.1, which is a number of atoms thick about 100nm).*

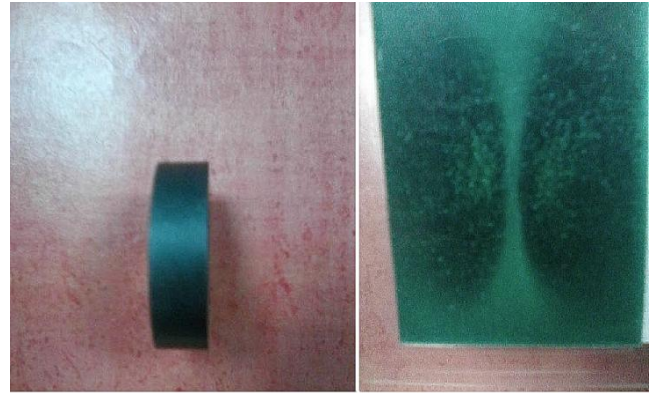

**Fig. 2. Bloch wall region of ferrite ring magnet (side view) as shown by a magnetic field viewer, at the middle of magnet as a light green (no magnetism) strip, dark areas (magnetism) left and right are the two poles of the axial magnetized magnet.**

On the other hand the ferrolens is a modern nanomagnetic photonic device that operates more at the quantum than the macroscopic level, has depth of field information and can therefore depict the quantum 3D image of the field of a magnet in real-time. As we will see in the next pages the actual quantum field [3] differentiates from the classic macroscopic iron filings image. We will analyze the experimental data, discuss and come to some surprisingly novel conclusions.

## 2. Materials and methods

As shown and described in our previous work [1], the ferrolens (i.e. commercially available under the registered trademark Ferrocell) was used as a direct observation nanomgnetic photonic device for the visualization of 3D magnetostatic fields in real-time (Video1 demonstration Link)[1] . Two optical grade glass disks are put together and sealed with optical cement around their periphery in a vacuum environment and with an encapsulated thin film placed in between the two disks. That is a 50 microns thick film of ferrofluid $Fe_3O_4$ yielding to a 10nm average size magnetite nanoparticles solution in a hydrocarbon based carrier fluid *(i.e. mineral oil)*. In addition the nanoparticles are coated with a surfactant *(i.e. oleic acid)*. Normally, ferrofuid in its free bulk state is opaque and blocks light. However, ferrofluid in a thin film configuration as described above, becomes transparent. Different types of neodymium magnets, such as cube, bar, cylindrical and ring magnets were placed under or above the ferrolens for observation. No need this time for the ferrolens to be fitted in a microscopy apparatus to view the magnetic fields. Different lighting conditions were applied from various artificial light sources but primary from an infrared remote controlled RGB LED light strip around the periphery of the ferrolens as shown in **fig.3**.

---

[1] Video1 demonstration link : https://tinyurl.com/yctntnjc

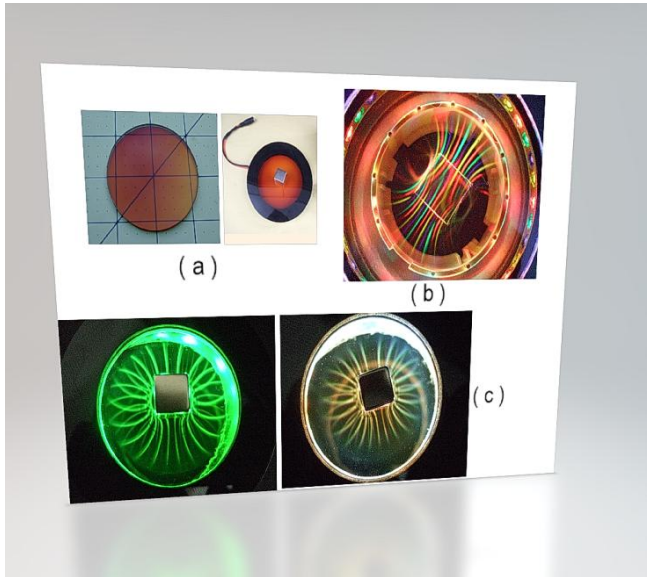

**Fig. 3.** (a) The ferrocell (left) and the ferrocell fitted in a 3D printed frame (right) with the LED light strip inside on the periphery. Cube magnet placed on top. (b) Cube neodymium magnet placed and in contact under the activated ferrocell. Notice under magnetic viewing the body of the magnet becomes transparent (i.e. invisibility cloak) and only its magnetic flux is shown. (c) Cube magnet placed this time on top of activated ferrocell, pole of magnet facing down, with different lighting configurations from the IR programmable RGB LED lighting strip. Green light (left) and white light (right). Depth of field information is shown.

A special geometry magnetic ring array prototype was also constructed for the purpose of our experiments. The plastic frame for fitting in the twelve, 1 mm thick, 10 mm square magnets used in this magnetic array design, was made using a 3D printer. This specific magnetic ring array, emulates the vortex-toroid geometry of the static magnetic field of a magnet shown by the ferrolens. A 3-axis xyz magnetometer was also used in the experiments for the measurement of the magnetic field strength in 3D space as shown in **fig.4**. Details about the design of the array as well as the 3-axis magnetometer will be described in our next publication.

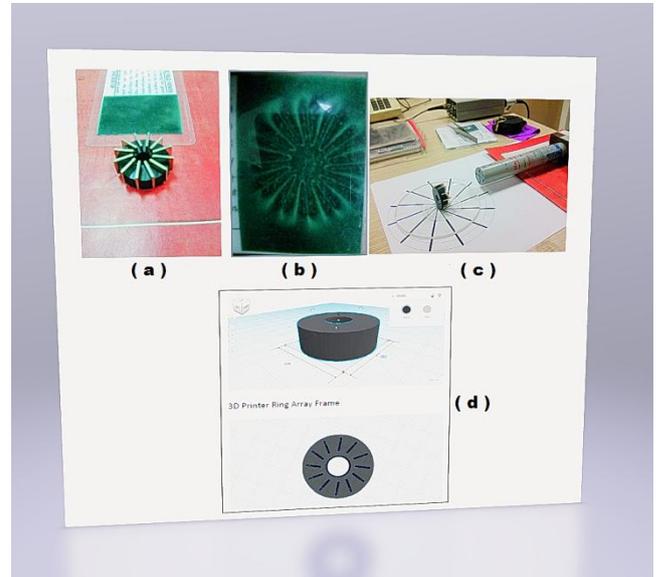

**Fig. 4.** (a) Magnetic ring array designed to emulate 3D geometry of magnetostatic fields observed by the ferrolens. (b) Pole of the magnetic ring array as shown by a magnetic viewer. (c) 3D field measuremets with a 3-axis xyz Mag-03MCESL70 magnetometer. (d) Magnetic ring array frame, 3D printer blueprint. We see the 12 slots where the individual magnets of the ring are placed in. This particular ring geometry placement of the magnets with skew angles emulates the field geometry of a magnet as observed with the ferrolens.

Various 2D&3D graphing, plotting and analysis software was used for some of the results of the research we present as well as other graphics packages, (x,y,z) position digitizer and 3D graphics illustration software.

## 3. Results

A cube neodymium magnet in fig.5 is placed very close under the ferrolens. The cube magnet is placed on its side with its two magnetic poles facing left and right as shown by the ferrolens as two dark circles. The ferrolens is lit by a programmable remote controlled RGB LED lighting strip on the periphery of the lens, programmed to emit white light (i.e. the natural color of the ferrofluid thin film is orange-brown). The result is all surfaces of the ferrolens to be lighted uniformly by the omnidirectional artificial light source. The superparamagnetic single domain [4,5] nanoparticles inside the ferrolens align with the external magnetic field of the magnet induced in the ferrolens following its magnetic flux and at the same time reflect part of the light therefore allowing them to *'paint'* the magnetic flux lines of the field and make them visible [1] (Video2 demonstration Link)[2].

---

[2] Video2 demonstration link: https://tinyurl.com/y78mgd7a

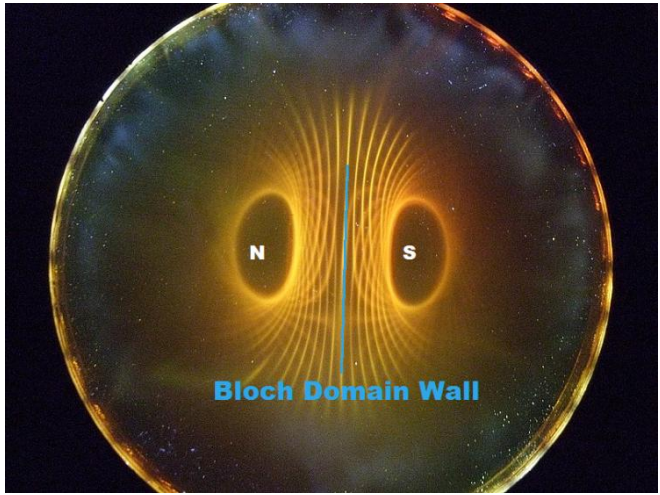

**Fig. 5.** Quantum field of a magnet shown by a ferrolens as two geometrical vortices, each at either pole of the magnet (black circles), oriented back to back and touching at the middle of the magnet where the Bloch domain wall ground state of the magnet is located (blue line in the middle). A strong cube magnet is placed under the ferrolens at a small distance (1-2 mm) therefore its body becomes effectively invisible and not shown by the ferrolens. Only its field image is projected. Superimposed text was used to indicate North and South Pole of the magnet and its Bloch domain wall. The photo is doing injustice in showing the real depth of field information actually displayed by the ferrolens. This information is difficult to fully capture in a 2-D photograph.

In the above fig.5 photograph taken from the ferrolens of the magnet's field, we can clearly see a compressed view (i.e. at the X-axis) emerging, of two vortices located spatially back to back and joined at the Bloch domain wall (ground state) of the magnet's field. **This is the quantum field of a magnet**. What is happening here? Is this true? How can flux lines are going straight through the Bloch region (i.e. region around Bloch domain wall, middle of magnet) of a magnet and in parallel to the Bloch wall axis (i.e. blue line)? All these will be explained at the discussion section of this paper. Also notice in the above photograph taken of fig.5 of the field, trajectories of the flux lines imply a counter geometry observed in the two mentioned vortices. North Pole appears to have counter clockwise rotation geometry whereas the South pole a clockwise.

**We must stress here**, that the first impression of interlacing (i.e. crisscrossing) flux lines appearing in photograph **fig.5**, of the magnetic field image on the ferrolens, are not actually crossing lines which would imply crossing of the force vectors tangent to the flux lines of the field. This would be of course an impossible and unacceptable condition, but actually **are overlapping lines of the field in 3D space which are shown by this ferrolens photograph as a compressed 2D image representation of the actual Euclidean magnetic field in space.**

**Nevertheless,** although 2D compression effect is inherent and amplified by the photographic lens, at the same time the ferrolens can depict a decent amount of **depth of field** information. Thus, in the actual viewing with the ferrolens, the observer will see one set of lines above the other, overlapping, as a **hologram.** This information is of course is impossible to be recorded by normal photography of the field shown by the ferrolens. Additionally and important, almost half of the lines you see are actually the *mirror image* of the pole lines projected to the other pole due the fact that the ferrolens is totally transparent to the magnetic field.

Fig.6(a) is the same photo of fig.5 without the text showing the side view of the quantum field of cube magnet. In figures 6 (b), (c) and (d) a very bright LED white light strip is used with a very strong bar magnet. These figures are very important and reveal the true geometry of the field at the **poles** of a magnet and the flux trajectories. As we said before, a ferrolens is totally transparent to the magnetic field induced therefore for a strong magnet in close proximity to the ferrolens and when oriented with its N-S pole magnetization axis perpendicular to the ferrolens surface, then, the field flux on its two poles, will appear on the ferrolens simultaneously and fully interlaced (i.e. criss-crossing lines) as a **2D compressed image representation of the actual 3D Euclidean magnetic field in space. This is exactly the case in photograph fig.6(d) where the bar magnet is placed on top of the ferrolens**. Again, actual observation with the ferrolens will result to holographic images and not to flattened 2D information shown in these photographs.

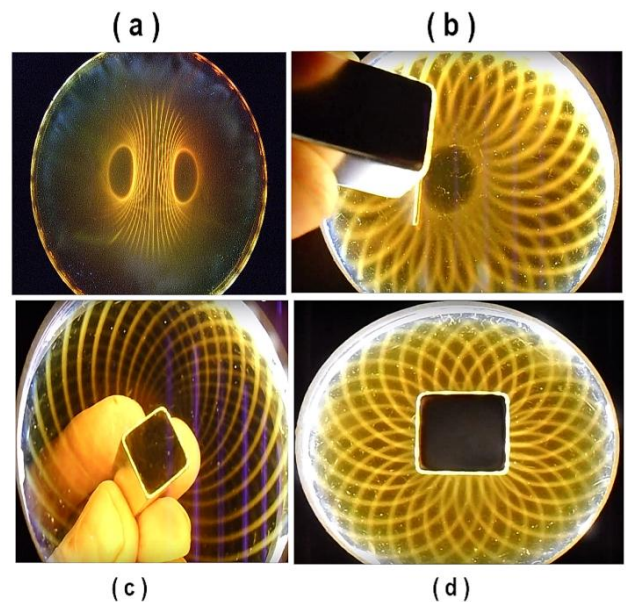

**Fig. 6.** (a) Previous photo in fig.5 without the text. (b) Strong bar neodymium magnet hold at a distance with its pole facing down to the ferrolens. Toroid geometry of quantum field of pole revealed without interlacing with its other pole flux. (c) Same magnet hold at a distance under the ferrolens some interlacing, criss-crossing, occurs. (d) Bar magnet placed on top of ferrolens with its pole facing down. Toroid fields of both poles of magnet appear now fully interlaced on the ferrolens.

**Specially**, as we see in **fig.6(b)** the pole of the bar magnet facing down to the ferrolens, is kept at a safe distance so that its field flux geometry can be clearly displayed by the ferrolens without the interlacing effect occurring this time (i.e. criss-crossing lines) with the flux of its opposite pole.

**Therefore,** the actual quantum field geometry of a single pole of a magnet and its flux trajectories is best demonstrated in fig.6(b). Notice how it appears like a rolled-in slinky.

In **fig.7(a)** a strong neodymium ring magnet is placed on top of the ferrolens with its pole facing down. Looking though its hole we see again the same quantum field appearing in the ferrolens proving the that field of all dipole magnets are the same and independent of their physical shape. Again because of the strong magnet used and due to its reduced height (i.e. 5 mm), the fields on both of its two poles appear together as a *compressed 3D image* with the flux lines of both of the poles interlaced and overlapping.

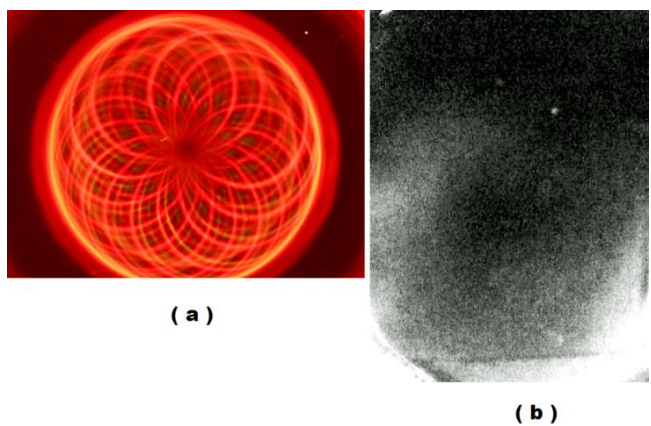

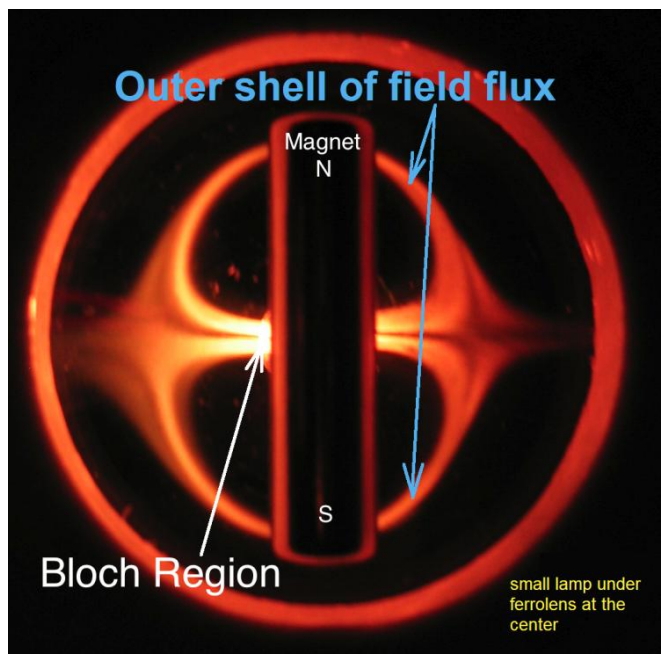

**Fig. 7.** (a) Quantum field of the poles of a ring magnet as shown by the ferrolens through the hole of the ring. The ferrolens LED light strip is emitting orange colored light. The ring magnet is placed on top of ferrolens with its pole facing down. Once more we see the same familiar pattern of the toroid fields of the two poles shown as a compressed 3D image fully interlaced and overlapping. (b) B/W photo for increased contrast, of a single pole of the field of the ring magnet shown under the ferreolens as a solid 3D toroidal shaped object. A black ferromagnetic painted paper was inserted between the ferrolens and the ring magnet**.**

We placed on top of the ring magnet in fig.7(b) a ferromagnetic paint black paper and then observed through a ferrolens from above (i.e. we kept the ferrolens a few cm away from the ring), so that the field of the pole of the ring magnet would emerge as a solid 3D shape this time. As we can see the field of the pole of the ring magnet is shown here by the ferrolens as a solid 3D toroidal object.

**Nevertheless,** what we consider the most important photographic evidence of the quantum field 3D geometry of any dipole magnet is what we see next.

At **fig.8** the total quantum field outline geometry of a dipole magnet is revealed here by the ferrolens. We can see clearly in the inner part of the photo, that the field consists of two separate and distinct magnetic flux bubbles or hemispheres, each at either pole of the magnet placed back to back and almost *'touching'* at the middle of the magnet where the Bloch region of the field is located separating the two. The experiment was contacted as explained in the fig.8 legend.

**Fig. 8.** The total quantum field outline geometry of a dipole magnet is revealed here by the ferrolens. We can see clearly in the inner part of the photo that the field consists of two separate and distinct magnetic flux bubbles hemispheres, each at either pole of the magnet placed back to back and almost tangent at the middle of the magnet where the Bloch region of the field is located separating the two. On top of the ferrolens a cylindrical magnet is placed as shown in the photo. A small incandescent lamp with a diameter smaller than the diameter of the magnet was placed directly under the ferrolens and almost in contact with it at the center. Light from the small lamp because its very close proximity to the cylindrical magnet, is mostly blocked by the magnet's mass and is strongly scattered sideways to the periphery of the lens revealing thereby the outline of the magnet's quantum dipole field. This photograph was taken with the aid of a custom-made mechanical servo apparatus fitted with a ferrolens, namely a **fluxscope**[3] as we call it. Notice here that the light outer ring is the outer rim of the lens and has nothing to do with the magnetic field shown inside. Notice the resemblance of the field shown with the Greek letter, theta θ.

This novel and groundbreaking observation concerning the geometry of magnetostatic dipole fields [6] and its importance, we will discuss later in this paper. The above photographic evidence also resembles the same image as when we look inside an apple cut-in half structure. We cannot dismiss the striking resemblance with the Greek letter theta, θ (a version of fig.8 without the overlaid text can be found is this link[4] ).

Of course the field depicted in the photo of fig.8 is just one shell of the actual quantum field of a dipole magnet. In reality there many overlaying repeating shells or layers, resembling an onion. This repeating pattern continues outwards all the way to the outer regions of influence a magnet exerts in 3D space. The magnetic flux density (i.e. magnetic dipole field strength) diminishes with distance using the inverse cube law.

---

[3] Fluxscope consists of two linear tracking mechanisms from CD players on a microscope stand with a motor control box. It can focus the ferrolens and light source distances between using switches and buttons. And control light brightness. It was designed for one light source (incandescent). Photo: https://tinyurl.com/ycdyfher

[4] Raw photo of fig.8: https://tinyurl.com/ycnrqgrw

As a result of **fig.8** we can say that at the quantum level of operation of a magnet as shown by the nanomagnetic superparamagnetic (i.e. single domain particles) ferrolens in contrast with the macroscopic iron filings experiment, the magnetic flux of a dipole magnet **does not consist of a single flux circuit**, closed between North and South poles but has two distinct and separate flux circuits. Each circuit closing between each pole and the middle ground state of a magnet where the Bloch domain wall region of the magnet is located. Furthermore, due the opposite spatial orientation of the two poles of a dipole magnet, these two distinct magnetic flux circuits must exhibit counteractive behavior. Although this is evident almost in all so far photos, a better demonstration of this effect is shown in **fig.9**. We will *discuss* all these in depth at the next section of the paper.

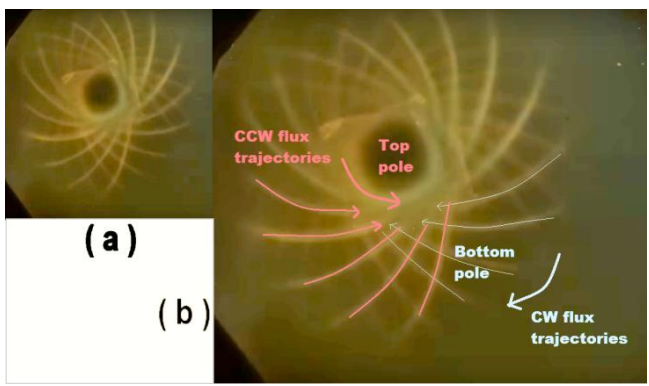

**Fig. 9. Small cube magnet hold under a dimmed light ferrolens so that few as possible flux lines appear with its top pole facing the ferrolens from below . Flux lines from both poles top and bottom pole of magnet are projected into the ferrolens, with the top pole flux trajectories (red color) overlapping in 3D space the more fade bottom pole flux lines (blue color) in an interlaced pattern, evidently demonstrating the counter geometry of the flux trajectories on the two poles of a magnet. Top actual pole shown is the North Pole of the magnet and bottom pole the South Pole. CCW here indicates counter clockwise and CW clockwise. Fig. 9 (a) is the original photo taken without the overlaid text and graphics in fig. 9 (b).**

Suffice to say here that in the in fig.9 we don't try to analyze or describe any flow and direction of energy namely from North to South pole of a magnet's quantum field, but merely the geometry of the flux lines in 3D Euclidian space and demonstrate their **actual counter geometry** on the two poles of a magnet.

The results of the measurements taken with the 3-axis magnetometer and their analysis will be discussed in the next section of our paper.

## 4. Discussion

In this section our primary focus will be on the obtained data from the experiments, which we will be *thoroughly* analyze and examine, for their validity and reliability. Also further analysis and data of the operation parameters of the ferrolens is provided in order to draw our conclusions.

Before the discussion turns into a debate whether the classical iron filings experiment or the ferrolens depicts the actual field of the magnet correctly or not, it is essential to say that both are correct in their display when their level of operation is considered.

The iron filings technique operates more at the macroscopic level showing us a macroscopic 2D imprint of the magnetostatic field of a permanent magnet dictated by the very low magnetic reluctance of the ferromagnetic iron filings *operating actually more like compass needles*. Thus, aligning only to the highest potential flux lines [3] towards the two poles of the magnet and neglecting the lower potential magnetic flux of the field of a magnet. Specifically, at the middle region of a magnet (i.e. Bloch region), the region with diminishing magnetism.

This effect is best demonstrated in **fig.10** bellow,

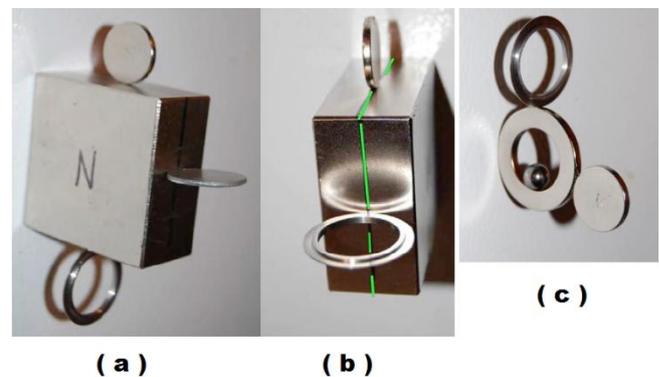

**Fig. 10. Magnetic attraction vs. ferromagnetic attraction difference demonstrated. (a) Small disk magnet is placed on top of a bar magnet side by side. Magnets get attracted and align themselves always when in this configuration, at their exact middle region where the Bloch domain wall axis is located and with both of the Bloch domain walls axes of the two magnets spatially coinciding and in <u>parallel</u> to each other. In contrast, iron disk and iron ring shown on the other sides of the block magnet get attracted by the magnet but always align and orient themselves to the path of minimum magnetic reluctance thus to the direction of the poles magnetization axis N-S of the magnet and <u>perpendicular</u> to the Bloch domain wall axis of the magnet. Notice also South Pole of small disk magnet on top is attracted by North Pole of bar magnet. (b) Bloch domain wall axis of the magnet indicated with a green line. (c) Same as in fig.10(a) but this time a ring magnet together with a disc magnet and a small sphere magnet was used. As before iron ring on top gets attracted and orients itself perpendicular to the Bloch domain wall axis of magnets.**

This inherent *limitation* of the ferromagnetic iron filings to align with the Bloch domain region of a magnet as demonstrated in **fig.10,** thus, they always align with strongest magnetic potential directions namely the two poles of a magnet, is the main reason why they fail to show any flux lines entering the Bloch domain region of a magnet. Therefore they fail to fully depict the quantum ground state field.

**On the other hand**, the superparamagnetic single domain magnetite ($Fe_3O_4$) nanoparticles inside the nanomagnetic ferrolens as demonstrated in the previous pages don't have these kinds of limitations iron filings have. Because they are single domain superparamagnetic [4,5] in nature and their high sensitivity, they can align with any flux line trajectory of the field of a magnet. Therefore, they're operating more on the quantum level thus able to depict accurately the quantum field of a magnet. The nanoparticles inside the ferrolens actually behave more like the small disk magnet on top of the bar magnet shown in fig.10(a) and not as the iron filings emulated by the iron ring in the same above figure.

Moreover, as shown in our previous research [1], these nanomagnetic particles in a thin film ferrofluid configuration while encapsulated inside the ferrolens are not subject of Brownian motion. As shown in our pervious paper [1] strong Van der Waals forces [7] essentially nullify Brownian motion and the nanoparticles inside the ferrofluid carrier are hold in a state of equilibrium.

The encapsulated thin film of ferrofluid inside the ferrolens in this state, does not flow, but exists in a balanced state of equilibrium no matter what position the cell is oriented. The nanoparticles inside the ferrolens do not settle with gravity. More in detail, the anionic surfactant coating [8] on the nanoparticles keeps the particles from touching each other (i.e. clumping or agglomeration) in the free state when there is no external magnetic field present. Notice here that the generated Van der Waals forces in the ferrofluid are not attractive but due to steric repulsion [9] results to stabilization.

Therefore, the nanoparticles movement is essentially we can say is dictated only by their induced by an external magnetic field magnetic moment according to their Néel relaxation time calculated [1,9,10] from equation (1),

$$\tau_N = \frac{1}{f_o} e^{\frac{KV_N}{KT}} \quad Néel\ relaxation\ time \quad (1)$$

$$V_N = \frac{4}{3}\pi R^3 \quad Néel\ particle\ size \quad (2)$$

for which $f_o$, and $K$ are the frequency constant of Néel relaxation (Larmor frequency), and the anisotropy constant of the particle, respectively and T the temperature in Kelvin units. Whereas $V_N$ is the Néel particle volume size and is given by (2) equation, where $R = d/2$ is the magnetic particle radius.

Although Néel relaxation time is more important for the response time of the ferrolens for dynamic magnetic fields, it still plays a significant role when the ferrolens is used in magnetostatics research and applications where real-time response is always desirable. Such, as for example, experiments which involve moving magnets or magnetic dipole interaction between magnets. In general, a small value of Néel relaxation time is needed in order for the ferrolens to display the information in real-time. As we have proven in our previous work the ferrolens with the 10nm particles can respond in real-time for dynamic magnetic fields or fast transient states up to 5MHz [1].

Concerning the optical and photonic properties of the ferrolens, the multiple different colored lines shown in some ferrolens configurations such for example in fig.3(b) is because a multiple colored RGB LED light strip source was used and these lines are *not product of interference* of light reflecting on the ferrolens surfaces. Color of lines is most dependent from light source color and tint slightly changes depending magnetic polarization of nanoparticles. As we have shown when white LED light is used, the result is lines to have same uniform color all over the ferrolens surface.

Optical **light refraction** index of the ferrolens is very small. Depending the carrier fluid (specially water based) used, can go **up to 90% transparency** as measured for all colors of light used in the ferrolens as demonstrated by the spectrograph in fig.11 below (all data measurements and excel graph can be found in this link[5] for download,).

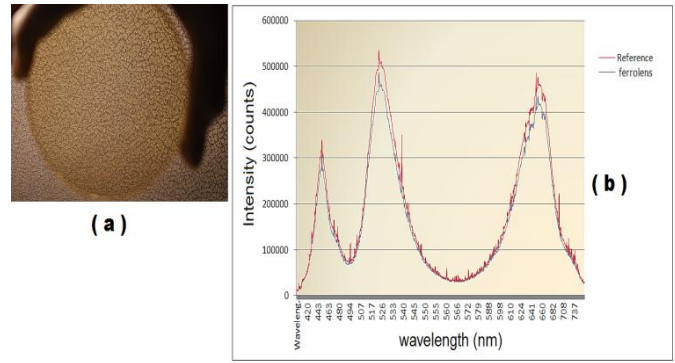

**Fig. 11. (a) Looking through an inactivated ferrolens (i.e. no external magnetic field is applied) to a pattern behind. (b) Spectrograph of a water based carrier fluid ferrolens through visible light spectrum. Red line is the reference spectrum obtained without the 50 μm thin film of ferrofluid inside the lens (i.e. two 2mm thick optical grade glass disks put together). Blue line represents the compete ferrolens with the encapsulated thin film of ferrofluid. As shown in the graph measured transparency did not drop below 90% at any point of the visible light spectrum.**

**Therefore, *light refraction index is controlled almost exclusively by magnetic polarization of ferrolens* [11,12] by the external magnetic field induced. Light dispersion effects [13] are also kept minimal due to the 50 microns or less transparent thin film and the two parallel optic-quality glasses used in the ferrolens which will mostly cancel out any small dispersion they may have.**

For the notion that the lines we see in the ferrolens could be perpendicular 90° to the actual magnetic flux lines at the xy 2D plane, **the answer** is that this is highly improbable to impossible to happen since we use uniform 360° omnidirectional lighting on the ferrolens and therefore if the above was the case that would totally mess up the display and no consistent geometrical pattern would be shown by the ferrolens. *In addition, the magnetic flux lines we see in the ferrolens would not end up at the physical locations where the poles of the magnet are (i.e. the two black holes depicted by the ferrolens).*

The theoretical argument also of if gyromagnetic precession *(i.e. Larmor frequency)* [3,14] calculated by equation (3) on this new observed by the ferrolens quantum field geometry, is maintained, can be positively answered since the new geometry observed by the ferrolens basically consists of two joined hemispheres making up a *sphere* (fig.8).

---

[5] Excel graph data of fig.11: https://tinyurl.com/y9dvxr9a

$$f = \frac{\gamma}{2\pi} B \quad \textit{Larmor frequency} \quad (3)$$

Where $\gamma$ is the gyromagnetic ratio and $B$ an external magnetic field. Notice in equation (3) the $2\pi$ factor essentially describing the circle, *remains unchanged* in the new observed geometry by the ferrolens of the quantum field of a magnetic dipole.

**Changing page,** the measurements taken with the **3-axis magnetometer** on the prototype special magnetic ring array constructed, shown previously in fig.4, emulating the toroid-vortex geometry of the quantum field of a magnet shown by the ferrolens specially at fig.6, 7(a) and 9, *confirmed* the *elliptical trajectory* of the flux lines on the poles of a magnet closing circuit between each pole and the Bloch domain region as demonstrated previously in fig.8. Also the geometrical rotation direction created by the skew angles of the magnetic flux lines trajectories on the poles was also confirmed, namely counter clockwise (CCW) trajectories on the North pole and clockwise for the South pole. The results are presented below in **fig.12** with surface maps measuring magnetic field strength around the poles 360° in a circle of the vortex geometry constructed magnetic ring array. Measurements were taken at 30° angular distance intervals around each pole.

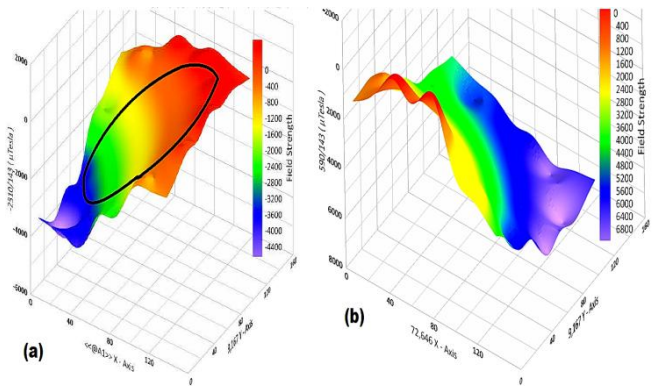

**Fig. 12.** Surface maps of magnetic field strength on the two poles of prototype special ring array emulating toroid-vortex structure of observed with the ferrolens quantum field of dipole magnets. Measurements were taken, with a 3-axis magnetometer at 30° angular intervals in a circle indicated with little humps or dips of the surface maps. Both maps have a slope evidently of the vortex geometry of the quantum field in the poles of a magnet. Also the counter behavior of the two poles is shown clearly by these surface maps. (a) **North Pole surface map**. A 2D projection of a single flux line ellipsoid trajectory (i.e. black ellipsoid) is drawn over the surface map by following the individual measurement points and the slope of the map, confirming therefore observations shown by the ferrolens previously of the flux lines geometry and trajectory on the poles of a magnet. Also the turned position of the map on X-axis indicates a CCW rotation geometry of the flux at the North Pole. (b) **South Pole surface map**. Same as before in fig. 12 (a) but this time the counter behavior of the two poles is evident. The position of the South Pole surface map indicates a CW rotational geometry of the flux lines trajectories on the South Pole of a magnet as shown by the ferrolens (see fig. 9b).

We mapped the surface of the prototype ring magnet, shown on fig.12 on the xy plane using a position digitizer. The Z-axis on the above maps of fig.12 indicates magnetic field strength measured in µTesla units. The µTesla values are calculated by dividing each value shown on the Z-axis in mV with the resolution of the used 3D-axis magnetometer thus, 143 mV per 1 µTesla. The individual measurements values taken with the 3-axis magnetometer can be found in this link[6].

To illustrate more clearly both flux trajectories happening on the different poles of a dipole magnet North and South as we have *observed with the ferrolens and confirmed with the magnetometer experiment*, the following graphical illustrations are presented in **fig.13(a)(b)** using two hyperboloids as shown. Two counter symmetrical flux trajectories are drawn sliding on the surfaces of the hyperboloids in fig.13(a), each for one pole on a magnet. Fig.13(b) also illustrates the counter geometrical rotation of the flux lines on the two different poles of the magnet North and South Pole. The joint area where the two hyperboloids meet in both illustrations fig.13(a)(b), represents the Bloch region of the quantum field of a magnet.

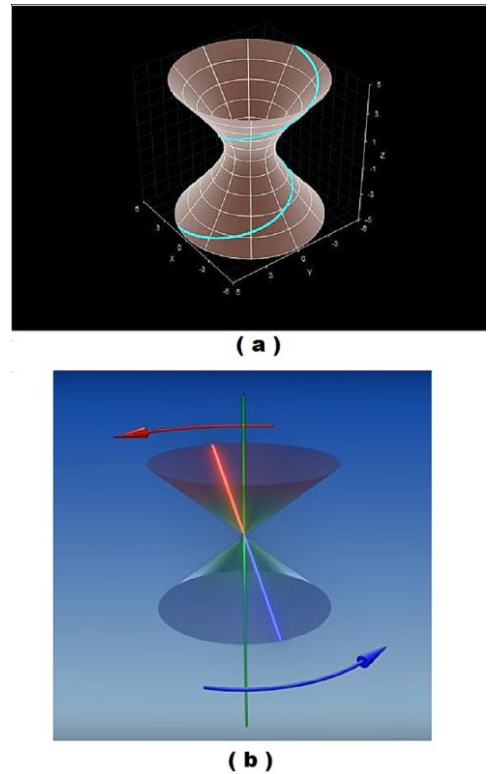

**Fig. 13.** (a) Graphical Illustration using two hyperboloids as shown, of the individual magnetic flux trajectories geometry on the two poles of a magnet, each hyperboloid representing one pole of the magnet. Bloch ground state region of the quantum field of a magnet is shown here as the joint area of the two hyperboloids. The counter directional behavior of the flux trajectories on the different poles of the magnet is apparent. (b) A second illustration indicating generally the counter geometrical rotation of the flux on the two poles of a magnet. Red arrow CCW rotation geometry for the North Pole and Blue arrow CW rotation for the South Pole of a magnet.

By graphical extrapolation of all the data mining we collected during this research with the ferrolens and magnetometer experiment and using graphical interpolation methods, the final graphical synthesized image of the vortex-toroid geometry of the quantum field of a dipole permanent magnet in *3D Euclidian* space is illustrated as below at fig.14.

---

[6] Three-axis magnetometer measurements: https://tinyurl.com/yapq8o8u

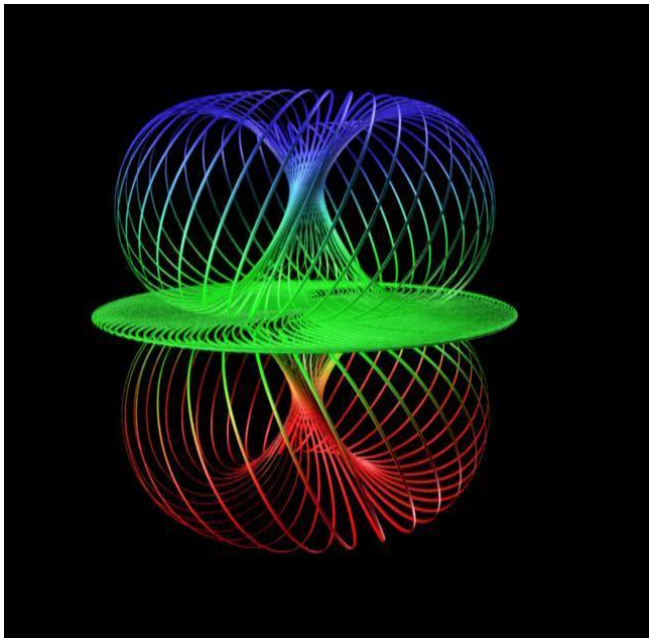

**Fig. 14. Quantum field geometry in 3D Euclidian space of a dipole magnet as observed with the ferrolens.**

## 5. Conclusions

Modern nanomagnetic direct observation devices for magnetic fields in general like the ferrolens or else known as ferrocell, give us the opportunity to observe more closely the magnetic field. Including the *depth of field* information of a magnet, which is actually considered a quantum mechanics device itself. Since magnetism falls more into and is described best today by quantum mechanics as a quantum effect than by electromagnetism general theory, a nanomagnetic real-time observation device would be the best choice for depicting the actual of what we call the quantum field of a magnet. As such, the nanomagnetic ferrolens using single domain (i.e. superparamagnetism) particles described herein, follows more precisely and in detail the magnetic flux of a magnet operating more at the quantum level and has not the limitations as we described and proved of the classical iron filings experiment which shows only a macroscopic imprint of the magnetic field, failing to depict quantum effects in the field of a magnet such as its Bloch domain wall region.

Novel observations of the field of magnet using the ferrolens were made, never seen before and with some surprising results. Although basically, the spherical geometry of the field of a magnet was confirmed, a closer examination at the Bloch region of the field of a magnet, made possible by the ferrolens, reveals that the actual geometry of the quantum field of a magnet consists of two separate magnetic bubbles toroid shaped, each around each pole of a dipole magnet placed back to back with both bubbles nearly tangent at the ground state Bloch region of the magnet (i.e. middle of magnet fig.8 & fig.14). ***Essentially the quantum magnetic field of a magnet consists of two hemispheres.*** Further observation of the individual flux lines trajectory inside these toroid fields we described, on the two poles of a dipole magnet North and South pole, revealed a *skewed ellipsoid trajectory geometry* [15] inside 3D Euclidian space for each magnetic flux line, closing circuit around each pole and the Bloch region. Additionally, this skewed flux creates elementary vortex geometry on the two distinct toroidal fields with *counter* rotational vectors (fig.13).

The above observations were confirmed with a 3-axis magnetometer on a prototype magnetic ring array which emulates this above described and observed with the ferrolens, complementary counter rotational toroid-vortex geometry of the quantum field of a dipole magnet.

Any argument of whether it is possible, the flux lines observed in the ferrolens to be products of light interference and other optical phenomena, were examined and proven invalid experimentally and theoretical and that light polarization in the ferrolens is exclusively controlled correspondingly, by the external magnetostatic field induced in the ferrolens.

Furthermore, gyromagnetic precession (i.e. Larmor frequency) is maintained in this new observed field geometry by the ferrolens since it basically consists of two joined hemispheres (fig.8).

We believe that our research presented on this paper here but also our previous work [1] with this new and exciting, modern version of the iron filings experiment, nanomagnetic direct observation device for magnetic fields in general called ferrolens, will be taken under serious consideration and study. That might propel research with the potential to lead to new breakthrough discoveries unveiling the true nature of magnetism.

## Acknowledgements

We like to thank Mr. Iraklis Rigakis, academic staff, for his assistance in the construction of the prototype magnetic ring array used in the experiments of this research. Also we like to thank Dr. John Chatzakis, academic staff for his support.